# Nix to the Rescue for a Reproducible HPC-AI Software Stack

Wenke DU[1,2] , Jean-Marc Gratien[2] , Raphael Gayno[2] , and Bruno Raffin[1]

## I. Introduction

Reproducibility in high-performance computing (HPC) critically depends on the ability to reconstruct complex software environments across machines and over time. In practice, this remains challenging. Production supercomputers typically operate under restrictive policies: no root access, limited or no internet connectivity, which constrain how software can be installed and deployed. The traditional solution relies on environment modules[1] managed by system administrators. While effective for well-established HPC stacks, this approach becomes increasingly brittle as software complexity grows: users often need packages beyond the precompiled modules provided, build configurations are opaque, and environments are hard to reproduce or port across systems.

Containers have emerged as a complementary approach and are now widely supported on HPC systems. They offer users greater control by encapsulating application environments into portable artifacts, partially decoupled from system constraints. However, containers alone do not guarantee reproducibility: constructing them often involves ad-hoc scripts or imperative workflows that are hard to version, audit, and adapt. As a result, rebuilding or modifying a container, such as updating dependencies, can be as fragile as traditional approaches.

These challenges are amplified in modern hybrid HPC/AI workflows. Consider coupling a numerical solver with a neural network component. The solver is typically written in C/C++ or Fortran and relies on MPI and optimized numerical libraries, while the neural network is often implemented in Python and accelerated on GPUs. The resulting software stack spans heterogeneous ecosystems with distinct tooling, dependency models, and compilation constraints. Ensuring consistency across these boundaries, while remaining compatible with system specific libraries for high performance networks and GPUs, pushes module-based and container-based approaches to their limits and raises significant reproducibility challenges.

Recent advances in package managers, such as Spack[2], Guix[3] and Nix[4], provide promising alternatives. These tools offer fine-grained, declarative control over software environments and are backed by large, actively maintained package collections. Spack, originating from the HPC community, emphasizes flexibility and interoperability with system-provided libraries, but this flexibility can weaken reproducibility guarantees. In contrast, Nix and Guix adopt a fully isolated approach, controlling the entire dependency graph down to low-level system libraries, thereby enabling stronger reproducibility over time. Importantly, both can generate container images directly from environment specifications, bridging the gap between environment definition and its deployment.

In this paper, we report on our experience locally building and remotely deploying a hybrid HPC/AI software stack under realistic supercomputing constraints. We begin with a conventional workflow based on modules and Conda[5], highlighting its limitations in terms of portability and reproducibility. We then demonstrate how Nix enables a fully reproducible environment specification, which can be used both for local development and to produce a container artifact successfully deployed on a production HPC system. Our results illustrate how declarative package management can simplify the construction, sharing, and long-term preservation of complex scientific software stacks.

## II. Modules, Conda and Containers

Our first approach combined Environment Modules with Conda to manage a mixed C++/Python project. For local development, Modules loads system libraries such as MPI and CUDA, while Conda handles the Python side and small utility C/C++ libraries. For remote deployment, we use Apptainer. Conda is installed inside it to recreate the same Python and C++ utility environments. The container is then transferred to the remote cluster. On the cluster, we `module load` the same system dependencies and bind-mount the relevant host directories into the image. Inside the image, the Conda environment supplies the remaining dependencies.

In practice, this combination needs significant manual intervention to work, which compromises reproducibility. First, Conda C/C++ packages do not follow a consistent installation layout that CMake can discover automatically. We had to manually set `CMAKE_PREFIX_PATH`, `FMT_ROOT`, `Torch_DIR`, and `LD_LIBRARY_PATH` to locate dependencies within `CONDA_PREFIX`. Second, Conda's environment isolation does not prevent CMake from finding system-installed packages outside the Conda environment. For system libraries like MPI and CUDA this is intended, but a library that happens to be present locally can be silently picked up by CMake. As a result, a locally "working" environment is not guaranteed to be reproducible on the cluster and debugging the mismatch is frustrating. Developing entirely inside the container would eliminate this leakage but complicates IDE integration significantly.

[1] Univ. Grenoble Alpes, Inria, CNRS, Grenoble INP, LIG, Grenoble, France, {wenke.du, bruno.raffin}@inria.fr.

[2] IFP Énergies-Nouvelles, Rueil-Malmaison, France, {wenke.du, jean-marc.gratien, raphael.gayno}@ifpen.fr.

More fundamentally, this workflow does not compose across projects. When a second project depends on the first and brings its own dependencies, it is unclear what the combined environment should even look like, and the developer is left to resolve both the version compatibility and the setup by hand. In some cases we ended up copying the source of smaller projects directly into a larger one for convenience, producing giant, non-modular codebases. This complicates environment setup further and amplifies the previous problems.

## III. Solution with Nix

Nix addresses these three problems directly. Its consistent package layout means dependencies declared in a Nix environment can be discovered automatically by CMake's `find_package`, with no need to set `CMAKE_PREFIX_PATH`, `*_ROOT`, or `*_DIR`; this reduces the project's `CMakeLists.txt` to simple `find_package` declarations. Its full environment isolation makes it impossible for unintended system libraries to leak in. And its flake system handles cross-project composition: each project declares the Git URLs of the projects it depends on as flake inputs and consumes them directly, reducing per-project environment complexity.

The same mechanism handles dependencies available only as prebuilt vendor binaries. We extract NVIDIA's PyTorch from their Docker image, supply standard dependencies from `Nixpkgs`, patch the ELF binaries with `autoPatchelf`, and republish as a standalone flake. Downstream projects consume it as any other Nix derivation.

Beyond fixing these three problems, Nix unifies C++ environments, Python environments, and container generation under a single tool. Python and C/C++ packages are declared together, and the same specification produces an Apptainer image with all runtime dependencies included, with no separate recipe required. Appendix A provides an example.

One inconvenience with Nix is that the development shell and the production build, defined by a Nix derivation, are separate environments, which forces application-specific build logic to be expressed twice. Both need the same set of build options but with different values. For example, debug flags in development and release flags in production. Maintaining two separate build invocations works but quickly becomes tangled. We resolve this with CMake presets, which aggregate the flags into a single definition with different values per preset.

Nix also has a relatively steep learning curve. Its functional language is unfamiliar to most computational scientists, and its packaging conventions take time to get familiar. We argue this cost is bounded: learning Nix is a one-time investment with predictable difficulty, whereas debugging ad-hoc environment errors is open-ended and recurring. In our experience, LLM assistance has further lowered this barrier by generating correct Nix files and helping diagnose Nix errors.

We have applied and validated this approach in three projects. The first is a framework for coupling parallel numerical solvers with neural networks. The second is a use case that consumes this framework alongside a prebuilt thermodynamic solver. The third is a patched Triton Inference Server PyTorch backend, linked with prebuilt NVIDIA PyTorch and itself consumed by the framework.

## IV. Discussion

Guix and Spack address similar problems but did not fit our setup. Our workstation has no root access. In the Nix community, `nix-user-chroot` allows installing and running Nix inside a Linux user namespace. Guix has recently gained support for running its daemon as an unprivileged user[6], but the installation step itself still requires root, and we did not find an equivalent user-space installer in the Guix community, so we chose Nix.

Spack addresses a different deployment model. Our workflow is local development followed by remote deployment, which requires full isolation from the host environment both during development and when generating containers. Spack primarily assumes the user develops, builds, and runs code in the same environment, so it uses the host toolchain and runtime by default. Achieving full isolation requires extra effort, and for container generation, the user must manually maintain `libc` compatibility between the host and the image, so we chose Nix over Spack.

Nixpkgs has the largest package coverage compared to Guix and Spack. Particularly, ML coverage is broad for mature packages but thinner for newer ones: `DeepXDE` is not yet packaged, and `onnxscript` currently lives only in the unstable channel. We work around via `buildPythonPackage` and pulling from multiple channels. Closing these gaps requires community-wide effort.

## Acknowledgement

Our positive experience with Nix is largely due to the quality of the `Nixpkgs` repository and the responsiveness of the Nix community. We are grateful for their effort in maintaining packages for the HPC and ML ecosystems.

## VI. Appendix A

### A. *Python and C/C++ dependencies declared together*

```
{
  inputs.nixpkgs.url = " ... ";

  outputs = { self, nixpkgs }:
    let
      sys = "x86_64-linux";
      pkgs = import nixpkgs { ... };

      py_env = pkgs.python313.withPackages (ps: [
        ps.numpy ps.pybind11 ps.torch
      ]);

      my_deps = with pkgs; [
        fmt cli11
        cudaPackages.cuda_nvcc
        cudaPackages.cuda_cudart
        py_env
      ];

      myProject = pkgs.stdenv.mkDerivation {
        # ...
        buildInputs = my_deps;
      };
    in {
      # Case 1: development shell
      devShells.${sys}.default = pkgs.mkShell {
        inputsFrom = [ myProject ];
      };

      # Case 2: Apptainer image
      # built from the derivation
      packages.${sys}.image =
        pkgs.singularity-tools.buildImage {
          name = "my-project";
          contents = [ myProject ];
        };
    };
}
```